\documentstyle[12pt]{article}
\newcommand{\squnu}{\sqrt{1 - v^2}}
\newcommand{\squTh}{\sqrt{1 +\theta^2}}

\begin{document}

\begin{center}
\huge
From the Field Electron Model to the Unified Field Theory
\end{center}
~\bigskip~
\section{Definition of Field Sources.}

Maxwell-Lorentz equations for potentials are taken as the basis.  Left parts 
of equations remain unchanged and are considered precise, but field sources 
in the right part a given a new definition.Spherically symmetrical electron 
at rest is described in the spherical coordinate system, $S$. There are two 
sets of currents with different charge signs in each point: 
$$
\vec{\jmath}_\pm(R,\psi) = \rho_\pm(R,\psi) \cdot \vec{v}(R,\psi), \quad
v_\pm(\infty, \psi) = 1, \quad
\lambda = \cos\psi = \frac{(\vec{v},\vec{R})}{v R},
\eqno{\rm(A)} 
$$
and the system of units in which the velocity of light is equal to one is 
selected.

These currents form stationary current threads of moving Ons ---
marked elements of continuos charged medium. Continuity equation is 
satisfied for each charge sign independently. The ends of each current 
thread rest on infinity and each thread has an opposite one, i.e. coinciding 
with the direct one to within velocity sign or time.

When coming from the 
infinity where the field is equal to zero, charged Ons are affected by the 
field of electron $A_e$ and form pairs of direct and opposite current threads in accordance with equation of On motion in the field

$$ J_a = D(A_e). $$

Current thread pairs generate field in correspondence with Maxwell-Lorents 
equations
$$ A_a = M(J_a),$$ 
and the sum of all fields generated by all current threads of two  charge 
signs gives us an electron field 
$$A_e = \sum_a A_a.$$

An electron is declared as a pair one element of which is the field, $A_e$, 
and the other element --- the whole set of current threads, $J_a$, of two 
charge signs. Each element of the pair gives birth to itself through its 
partner. 
$$
\left\{\begin{array}{lcl}
A_e & = & \sum_a M(D(A_e)) \\[2mm]
J_a & = & D (\sum_a M(J_a))
\end{array} \right.
\eqno{\rm(B)}
$$

Operator $M$ affecting $J_a$ gives the field coinciding in form with 
the integral of sum of retarded and advanced potentials of 
Lienar-Vichert.

But now, each of two summands is an average of an retarded 
field of an On of one current thread and an advanced field of an On of an 
opposite current thread of this pair. Both summands have equal rights and 
their division into retarded and advanced terms becomes arbitrary.

The natural character and non-eliminability of $M$ operator symmetry 
relative to time reversal is the strongest reason for accepted definition of 
field sources, the properties of which are in conformity with the necessity 
to make vanish the sum of currents in each point of electron at rest and are 
guaranteed by the invariance of the law of motion with respect to time 
reversal.

The law of motion is declared a gnosiological derivative of the 
fundamental fact of existence of identical electrons with all their 
properties of complete objects in the terms of Maxwell-Lorentz equations 
with accepted definitions of sources (A) and the system of equations (B).  

Formally, the law of motion $D$ is a possible solution of the system of 
equations, (B), if all values are stationary and spherically symmetrical, 
currents are uniform and isotropic at the infinity, and $A_e$  asymptotically 
approaches a Coulomb field for large $R$.

\section{The Law of Motion.}

An arbitrary motion may be approximated by tangent uniformly  accelerated 
motion when the acceleration is constant in an instant accompanying 
reference system $\Sigma$, as the simplest one after inertial motion.

Suppose, as a 
first approximation, that this symmetry is global and all currents in the 
whole space have the same $W_\Sigma$. This $W_\Sigma$  is made equal to one 
by selecting an appropriate system of units.

Three-dimensional equation of such motion
$$ 
(1 - v^2)\ddot{\vec{v}} + 3(\vec{v},\dot{\vec{v}})\dot{\vec{v}} = 0
$$ 
is transformed into the following system in the cylindrical coordinate 
system:
\begin{eqnarray*}
\vec{v}(t) & = & \{v_x, v_y, 0\} = 
  \{v \cos \alpha, v \sin \alpha, 0\}, \\
\dot{\alpha} & = & \chi \frac{\left(\squnu\right)^3}{v^2},
  \chi = {\rm const}, \\
\dot{v}^2 & = & \frac{\left(\squnu\right)^3}{v^2}
  ((1 + \chi^2)v^2 - \chi^2).
\end{eqnarray*}

The special solution is selected such that: \\
1)  for $t \to +\infty$ the velocity is parallel to $OX$; \\
2) minimum vertex distance $\vec{s}$ and vertex velocity $\vec{\beta}$ 
are simultaneous at $t=0$  and   orthogonal.

Then the constant is fixed
$$ 
\chi = \mp\beta\gamma, \quad \gamma = \frac{1}{\sqrt{1 - \beta^2}}
$$
and the solution takes the following form:
$$
\vec{r}(t) = ( A \mp \squTh)\frac{\vec s}{s} + t\vec{\beta}, \quad
\vec{v}(t) = \mp \frac{1}{\gamma} \frac{\theta}{\squTh} \frac{\vec s}{s}
+ \vec{\beta},
\eqno{\rm (G)}
$$
$$
\frac{\vec s}{s} = \left\{\mp\frac{1}{\gamma}, \beta, 0\right\}, \quad
\frac{\vec \beta}{\beta} = \left\{ \beta, \pm\frac{1}{\gamma}, 0\right\},
$$
$$
A = s \pm 1, \quad \theta = \frac{t}{\gamma}, \quad 
\Gamma(t) = \frac{1}{\squnu} = \gamma\squTh .
$$

Suppose that the upper sign in symbols $\pm$ and $\mp$ represents 
positively charged currents, and the lower one --- negatively charged. 
Vertex velocities $\beta(s)$ increase continuously from $\beta(1\mp1)=0$ to 
$\beta(\infty) = 1$. The domain of radius 2 in the origin is unapproachable 
for negatively charged currents.

Motion in accordance with (G) may be represented in the form:
$$
\vec{W} = \mp\frac{1 - v^2}{A\sqrt{1 + \theta^2} \mp1}
(\vec{r} - t\vec{v}), \quad
t = (\vec{r},\vec{v})\cdot
\left(1\pm \frac{A}{\gamma} 
\frac{1}{\gamma\squTh \mp \frac{A}{\gamma}}\right),
\eqno{\rm (C)}
$$
being similar to the expression:
$$
\vec{W} = \frac{\squnu}{\mu}
\left(\frac{d\vec{p}}{dt} - 
\left(\frac{d\vec{p}}{dt} + (\nabla\mu)\squnu,\vec{v}\right)
\vec{v}\right),
\eqno{\rm(D)}
$$
obtained from differentiated with respect to $t$ inequality
$$ \vec{p} = \frac{\mu}{\squnu}\vec{v}, $$
that coincides with the classic law of motion of a point charge in the field 
at $\mu = \rm const$.

The motion of an On in the field as of an infinitely small element  
of a current thread cannot depend of the value of its charge $\Delta q$, and 
mass and pulse in the following form may be assigned to it:
$$
\Delta\mu = |\Delta q|\cdot\mu(t), \quad
\Delta\vec{p} = \frac{\Delta\mu}{\squnu} \vec{v} .
$$

Applying the law of conservation of impulsive moment for the motion 
of an On along current thread in spinless approximation and requiring 
$$
\frac{\mu(+\infty)}{\squnu} = 1,
$$
independently of $y(+\infty)$, and selecting appropriate mass unit, we 
obtain:
$$
\mu(t) = \frac{1}{\gamma\squTh \mp \frac{\gamma}{A}} .
$$

For vertex velocities satisfying the condition
$$ \gamma = A = s \pm 1, \eqno{\rm (E)} $$                               
the whole expression is simplified:
$$
\Gamma = r \pm 1, \quad \Delta\mu = \frac{|\Delta q|}{r} .
$$

Expressions (C) and (D) become identical if:
$$
\frac{d\vec{p}}{dt} = \Delta q \vec{E}, \quad
\vec{E} = -\frac{1}{r(r\pm1)} \frac{\vec{r}}{r} ,
$$
that justifies hyperbolic current field for large $r$.

Asymptotically similar expression may be obtained from the Lagrangian
$$
L = -|\Delta q|[\varphi - \vec{v}\cdot\vec{A}]_\Sigma -
\Delta q[\varphi - \vec{v}\cdot\vec{A}]_S ,
\eqno{\rm(L)}
$$
which is obtained from classic using substitution
$$ m_0 \to |\Delta q| \varphi . $$

The first terms is taken in instant accompanying reference system 
$\Sigma$, and the second --- in the system of inertia center $S$. Then, it is 
necessary to make a substitution
$$ \varphi \,\to\, \varphi - (\vec{v},\vec{A}) $$
in each term of (L) and proceed to density.

\section{Field Generated by Currents.}

Application of continuity equation to current field (G) gives us 
charge density along current thread:
$$
\rho(t,s) \squnu = 
1 \pm \frac{\gamma}{A}\, \frac{1}{\Gamma \mp \frac{\gamma}{A} + t} ,
$$ 
where normalization is carried out securing independence of this expression  
of $y(+\infty)$ at the infinity, and corresponding charge unit is selected. 
Namely this expression
$$ \rho_S = \rho(t,s)\squnu $$ 
is the charge density in the system of electron center of mass $S$, but this 
question is worth separate consideration.

For vertex velocities (E), differential charge density is equal:
$$
d\rho_S(r,\lambda) = \frac{1}{r}
\left(1 \pm \frac{1}{\Gamma}\,\frac{1}{1+v\lambda}\right)
\Gamma d(v\lambda).
$$
  
Scalar potential at the distance $R$ from the center of electron generated 
by currents from the spherical layer of radius $r$ with the center in the 
origin, omitting elementary integration with respect to angle variable, 
which currents are not dependent of, and a numerical factor, is equal to:
$$
d\varphi = \frac{1}{R}
\left(\int_\lambda J(\alpha,v,\lambda) d\rho_S(r,\lambda)\right)
4\pi r^2 dr,
$$ 
$$
J(\alpha,v,\lambda) = \frac{1}{4}
\int_{-1}^{1} \left(\frac{1}{\sqrt{A_+}} + \frac{1}{\sqrt{A_-}}\right)
d\sigma,
\eqno{\rm(J)}
$$
$$
A_\pm = B^2_\pm - v^2(1 - \lambda^2)(1 - \sigma^2), \quad
B_\pm = \sqrt{1 + \alpha^2 + 2\alpha\sigma} \pm v\lambda(\sigma + \alpha),
\quad \alpha = \frac{r}{R}.
$$

For $r \leq R$, the expression (J) is degenerates into elementary integral
$$
J(\alpha \leq 1, v, \lambda) = \frac{1}{2v}
\ln\left(\frac{1 + v}{1 - v}\right), \quad
v = v(r, \lambda),
$$
which becomes equal to one at $v = 0$, that was the reason 
why unit charge constant was omitted in (J). For inner points, when 
$R < r$, (J) 
is substantially elliptical and the following integral inequality is 
satisfied:
$$
\alpha \int_0^1 J(\alpha > 1, v ,\lambda) d\lambda =
\frac{1}{2v} \ln\left(\frac{1+v}{1-v}\right) .
\eqno{\rm (II)}
$$

Unexpectedly, reverse field has appeared in the inner volume when 
$0 \leq R < r$  with the sign of intensity opposite to intensity of direct 
field in the   outer domain, when $r \leq R <\infty$. 

The intensity of electron field is he sum of intensities of direct and
reverse fields of currents from spherical layers covering the whole space.

Vertex velocities (E) do not give needed field.

Approximate calculations of vertex velocities defined by the condition
$$
\gamma = \sqrt{s^2 - 1 \pm 2} ,
\eqno{\rm(E2)}
$$
of field intensity near zero result in the function close to linear: 
$$ E(R \ll 1) = {\rm const}\cdot R $$
and more realistic field for $R \gg 1$.

Consideration of the spin when applying the law of conservation of On
moment  for motion along current thread in the hyperbolic approximation  for
currents  changes the mass at  rest $\Delta\mu$ making  it less singular 
and accordingly 
decreasing the intensity of  field necessary to generate  On currents in the
vicinity of the center. 
        
The  beauty of  Lagrangian (L)  requires    calculations of field and 
currents on its base.

Interpretation  of  the  hyperbolic  solution (G) has  been made  
assuming
that  the  integral of intensities of reversed fields generated by spherical
layers  from $R$ to $\infty$ exceeds  the  integral  from $1\mp1$ to $R$ 
for all $R$ by absolute
value.   This  represents  the  field  of  electron  similar to the field of
stationery  negative  charge  distributed  near the center as the residue of
prevailing  reverse  field of  positively charged currents.  Such concept of
electron field structure seems to be preferable. 

The model obtained by substitution: $(\pm) \to (\mp)$  not affecting 
signs of electron  field and its mass cannot be excluded yet. 

\appendix
\section*{Appendix.}

Expansion the integral (J) in an infinite series takes the folowing form:
\begin{eqnarray*}
J(\alpha > 1, v, \lambda) & = &
\frac{1}{\alpha P}\left\{ 1+\frac13\delta^2 +
\left(\frac15 +\frac Q3 p_{10}\right)\delta^4 +
\left(\frac17 +\frac Q5 p_{12}\right)\delta^6 + \right. \\
& & \left.\left(\frac19 +\frac Q7 p_{14} +\frac{Q^2}{5}p_{20}\right)
\delta^8 + \cdots \right\},
\end{eqnarray*}
$$
P = 1 - v^2\lambda^2b, \quad b = \frac{\alpha^2 - 1}{\alpha^2}, \quad
Q = 4(\alpha^2 - 1)\lambda^2(1 - \lambda^2), \quad
\delta^2 = \frac{v^2}{\alpha^2P} ,
\eqno{\rm (J0)}
$$
\begin{eqnarray*}
p_{10} & = & 1, \\
p_{12} & = & \frac{2\cdot 3}{1 \cdot 2}\left(
1 - \frac{2\cdot1}{1\cdot3}\lambda^2\right), \\
p_{14} & = & \frac{3\cdot 4}{1 \cdot 2}\left(
1 - \frac{2\cdot2}{1\cdot3}\lambda^2 + 
\frac{2\cdot4\cdot1}{1\cdot3\cdot5}\lambda^4\right), \\
p_{16} & = & \frac{4\cdot 5}{1 \cdot 2}\left(
1 - \frac{2\cdot3}{1\cdot3}\lambda^2 +
\frac{2\cdot4\cdot3}{1\cdot3\cdot5}\lambda^4 -
\frac{2\cdot4\cdot6\cdot1}{1\cdot3\cdot5\cdot7}\lambda^6\right),\\
\ldots & = & \ldots; \\[2mm]
p_{20} & = & 1,  \\
\ldots\ & = &  \ldots; \\[2mm]
\ldots\ & = &  \ldots \ .
\end{eqnarray*}

Differentials of intensities of direct and reverse fields generated
by currents from spherical layers take the folowing forms:
$$
(r \le R)\colon\quad d\vec{E}_\pm(R,r) =
\pm\frac{4\pi}{R^2} \int_\lambda \frac{1}{2v}
\ln\left(\frac{1+v}{1-v}\right)  rd\rho_Srdr\frac{\vec R}{R} ,
$$
$$
(R <r_*)\colon\quad d\vec{E}_\pm(R,r_*) =
\mp8\pi R \int_\lambda R(\alpha,v,\lambda)r_*d\rho_S
\frac{dr_*}{r_*^2}\frac{\vec R}{R} ,
$$
$$R(\alpha,v,\lambda) = A_{20}v^2 + 2(A_{40} - bA_{41})v^4 +
3(A_{60} -2bA_{61} + b^2A_{62})v^6 + \cdots \ ,
\eqno{\rm(J^*0)}
$$
\begin{eqnarray*}
A_{20} & = & \frac13 - \lambda^2 , \\
A_{40} & = & \frac15 - \lambda^2 + \frac{2\cdot1}{1\cdot3}\lambda^4, \\
A_{60} & = & \frac17 - \lambda^2 + \frac{2\cdot2}{1\cdot3}\lambda^4 -
\frac{2\cdot4\cdot1}{1\cdot3\cdot5}\lambda^6 , \\
A_{80} & = & \frac19 - \lambda^2 + \frac{2\cdot3}{1\cdot3}\lambda^4 -
\frac{2\cdot4\cdot3}{1\cdot3\cdot5}\lambda^6 +
\frac{2\cdot4\cdot6\cdot1}{1\cdot3\cdot5\cdot7}\lambda^8, \\[2mm]
\ldots & = & \ldots; \\[2mm]
\ldots & = & \ldots\ .
\end{eqnarray*}

\bigskip
\noindent
Riga, March-April 1998.
\hfill
Alexander Semyonovich Zazerskiy.

\end{document}